\documentclass[twocolumn,a4paper,superscriptaddress,floatfix,showpacs]{revtex4}
\usepackage{amsmath,amssymb,epsfig,color,textcase}

\begin{document}

\title{\emph{Ab initio} investigation of the exchange interactions in Bi$_2$Fe$_4$O$_9$: The 
Cairo pentagonal lattice compound}

\author{Z.V.~Pchelkina}
\email{pzv@ifmlrs.uran.ru}
\affiliation{Institute of Metal Physics, S.Kovalevskoy St. 18, 620990 Ekaterinburg, Russia}
\affiliation{Ural Federal University, Mira St. 19, 620002 Ekaterinburg, Russia}

\author{S.V.~Streltsov}
\email{streltsov.s@gmail.com}
\affiliation{Institute of Metal Physics, S.Kovalevskoy St. 18, 620990 Ekaterinburg, Russia}
\affiliation{Ural Federal University, Mira St. 19, 620002 Ekaterinburg, Russia}
\email{streltsov@imp.uran.ru}


\date{\today}

\begin{abstract}
We present the \emph{ab initio} calculation of the electronic structure and 
magnetic properties of Bi$_2$Fe$_4$O$_9$. 
This compound crystallizes in the orthorhombic crystal structure with the Fe$^{3+}$ ions forming the 
Cairo pentagonal lattice implying strong geometric frustration. The neutron diffraction 
measurements reveal nearly orthogonal magnetic configuration, which at first sight
is rather unexpected since it does not minimize the total energy of the pair of magnetic
ions coupled by the Heisenberg exchange interaction. 
Here we calculate the electronic structure and exchange integrals of Bi$_2$Fe$_4$O$_9$ 
within the LSDA+U method. 
We obtain three different in-plane ($J_3$=36 K, $J_4$=73 K, $J_5$=23 K) and two 
interplane ($J_1$=10~K, $J_2$=12 K) exchange parameters. The derived set of 
exchange integrals shows that the realistic description of Bi$_2$Fe$_4$O$_9$ needs 
a more complicated model than the ideal Cairo pentagonal lattice with only 
two exchange parameters in the plane. However, if one takes into account only two 
largest exchange integrals, then according to the ratio $x\equiv J_3$/$J_4$=0.49$<\sqrt{2}$ (a critical
parameter for the ideal Cairo pentagonal lattice, see. Ref.~1) the 
ground state should be the orthogonal magnetic configuration in agreement with experiment. 
The microscopic origin of different exchange interactions is also discussed. 
\end{abstract}

\maketitle

\section{Introduction \label{intro}}
Until D. Shechtman discover quasicrystals (the Nobel Prize in Chemistry in 2011) 
it was thought that it is impossible to pack atoms into a regular lattice and obey a 
pentagonal symmetry.~\cite{Shechtman84} This type of the symmetry is not rare in the 
nature. It can be found in wildflowers and many sea dwellers as well as in scale of 
a fir cone and a pineapple. The only known naturally occurring quasicrystal phase is 
the icosahedrite (Al$_{63}$Cu$_{24}$Fe$_{13}$) found in the Koryak Mountains in Russia. 
The quasicrystals reveal a new class of the organization of the matter
with regular but non-periodic lattice. Such patterns have been known in the mathematics 
since antiquity, and medieval Islamic artists made decorative, non-repeating 
tessellation (the Cairo pentagonal mosaic).

Such an exotic and rare structure is the subject of a keen interest from 
both experimental and theoretical point of view.  As the number of bonds per 
elemental ``brick'' in the pentagonal lattice is odd the nearest neighbor antiferromagnetic (AFM) 
interactions would lead to the geometrical frustration. At present the most studied 
2D magnetic frustrated lattice is the triangular one consisting of regular 
polygons with equal nearest neighbor exchange interactions. Contrary to the triangles 
it is impossible to fill the plane with regular pentagons, the ``bricks'' of another shape 
are needed like in the Penrose lattice. Such a tessellation however, can be constructed using 
non-regular pentagons as in the case of the Cairo pentagonal lattice. 

The comprehensive analytical and numerical investigation of the antiferromagnetic Heisenberg model on 
the Cairo pentagonal lattice have been recently presented.~\cite{Rousochatzakis12} 
A simple pentagonal lattice studied in Ref.~\onlinecite{Rousochatzakis12} consists of two 
inequivalent sites with three and four nearest neighbors 
(see Fig.~\ref{magn.str} and Fig.~1 in Ref.~\onlinecite{Rousochatzakis12}). 
It has two types of the nonequivalent 
bonds, which connect threefold sites with each other ($J_{33}$ exchange constant) 
and threefold sites with fourfold ones ($J_{43}$ exchange path). Such a Cairo pentagonal 
lattice has a square Bravais lattice and the unit cell containing four fourfold- and two 
threefold-coordinated sites. The phase diagram of the AFM Heisenberg model was obtained 
as a function of the ratio $x \equiv J_{43}/J_{33}$ and spin $S$. In the classical limit 
(large $S$) three magnetic phases have been found: 1) the phase, where
the spins on the neighboring sites are orthogonal to each other ($x<\sqrt{2}$), 
2) a collinear 1/3-ferrimagnetic phase ($x>2$) and 3) an intermediate mixed phase ($\sqrt{2}<x<2$) 
which is a combination of 1) and 2).\cite{Rousochatzakis12}

At present there are known only two complex iron oxides which represent 
the physical realization of magnetic Cairo pentagonal lattice, 
namely Bi$_2$Fe$_4$O$_6$~\cite{Rousochatzakis12} and Bi$_4$Fe$_5$O$_{13}$F.~\cite{Abakumov2013}
Bi$_2$Fe$_4$O$_6$ can be obtained as a by-product in the synthesis of the multiferroic BiFeO$_3$ 
and seems to reveal multiferroic properties by itself.~\cite{Singh08} It is also regarded 
as a perspective material for the semiconductor gas sensors.~\cite{Poghossian91}

\begin{figure}[b!]
 \centering
 \includegraphics[clip=false,width=0.35\textwidth]{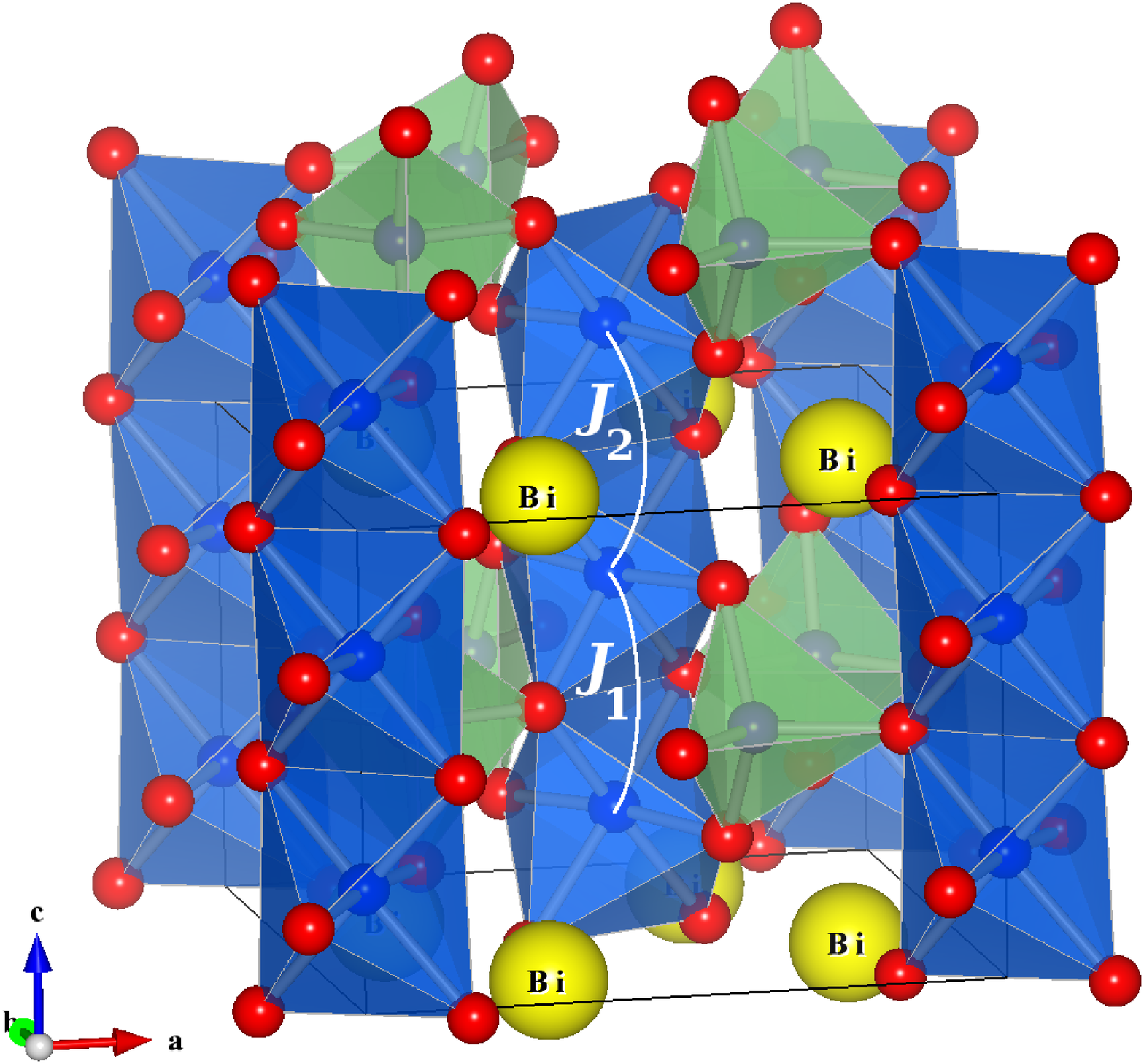}
 \includegraphics[clip=false,width=0.45\textwidth]{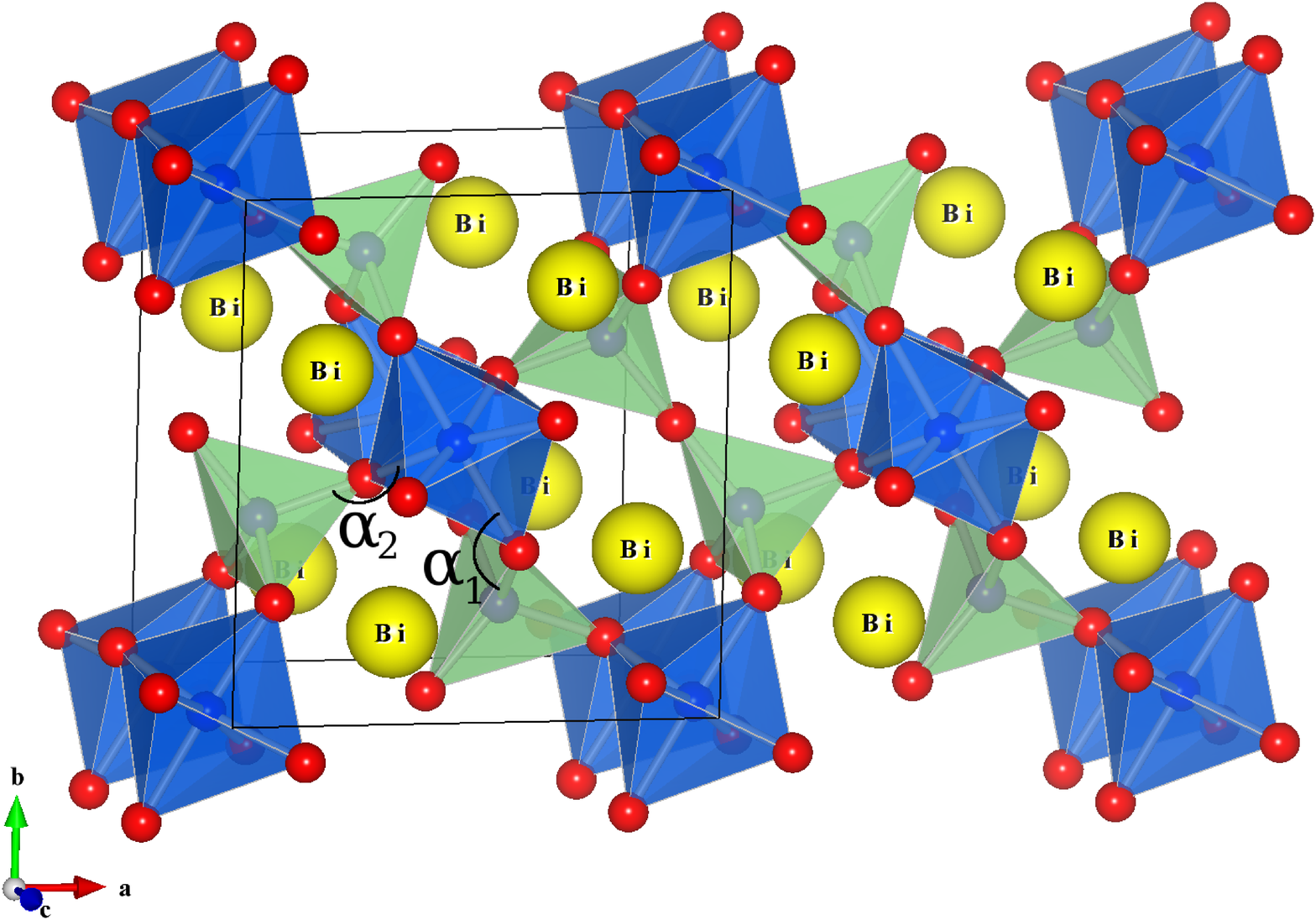}
\caption{\label{cryst.str}(color online). The crystal structure of 
Bi$_2$Fe$_4$O$_9$ along the $c$ axis (upper panel) and in the $ab$ plane
(lower panel). There are two types of the Fe ions:
Fe$_o$ is placed in the oxygen octahedra (blue), while Fe$_t$ is
in the ligand tetrahedra (green). O and Bi are shown as red and yellow balls
respectively. $J_1$ and $J_2$ are the interplane exchange interactions. 
We use VESTA software~\cite{MommaK.Izumi2011} for visualization.}
\end{figure}

Bi$_2$Fe$_4$O$_6$ crystallizes in a complex orthorhombic structure~\cite{Tutov1964, Ressouche-09} with 
the space group P\emph {bam} (No. 55). It has two formula units in the unit cell and 
two nonequivalent iron atoms Fe$_t$ and Fe$_o$ occupying the tetrahedral and octahedral 
positions, correspondingly (see Fig.~\ref{cryst.str}). The edge-sharing Fe$_o$O$_6$ octahedra 
form chains along the \emph c direction and these chains are bind by 
the corner-sharing Fe$_t$O$_4$ tetrahedra and Bi atoms.
Fe$_t$ occupies aforementioned threefold-coordinated sites, while Fe$_o$ - fourfold. 

The magnetic measurements on the single crystals were performed in Ref.~\onlinecite{Ressouche-09}. 
At high temperature a Curie-Weiss fitting of the magnetic susceptibility gives the 
paramagnetic temperature $\theta_p \approx$-1670 K and the effective magnetic moment 
$\mu_{eff}$=6.3(3)$\mu_B$ per iron atom in agreement with value 5.9$\mu_B$ corresponding 
to $S$=5/2 of the Fe$^{3+}$ ions. The long range magnetic order with 
T$_N$=238 K sets in at much lower temperature 
indicating the presence of magnetic frustrations in the systems. In the unusual 
nearly orthogonal magnetic structure  
at low temperatures the moments on all the iron atoms lying in the (\emph {a,b}) plane. 
The Fe$_o$ spins form four orthogonal sublattices while 
the Fe$_t$ spins align antiferromagnetically with each other 
(see Fig.~\ref{magn.str}). 

In contrast to the perfect Cairo lattice model studied in 
Ref.~\onlinecite{Rousochatzakis12} the real orthorhombic crystal structure of 
Bi$_2$Fe$_4$O$_6$ has few distinct features. Namely, each pentagonal unit cell contains 
seven sites because there are two Fe$_o$ ions in the center of the unit cell 
(Fig.~\ref{cryst.str}) with different coordinate along the \emph c axis,   
while the ideal structure has only one. Hence, for the realistic treatment of the 
magnetic interactions one needs to calculate at least five different
exchange constants, which can hardly be done reliably by fitting
the model solutions of the Heisenberg Hamiltonian to the magnetic 
susceptibility or other experimental observables. Such a fitting
allows however to estimate the ratio between some of the exchange
integrals.~\cite{Ressouche-09}

In this paper we present the \emph {ab initio} calculation of the exchange 
constants in Bi$_2$Fe$_4$O$_6$, compare the result obtained with available
theoretical and experimental data, show that this system
cannot be considered as a realization of the perfect Cairo pentagonal lattice, 
and discuss the microscopic mechanisms, which define the strength of different 
magnetic interactions.


\section{Calculation details}
We used the linearized muffin-tin orbitals method (LMTO)~\cite{Andersen1984} with the 
von Barth-Hedin version of the exchange correlation potential~\cite{Barth1972}
to calculate the electronic and magnetic properties of Bi$_2$Fe$_4$O$_9$.
In order to take into account strong electronic correlations on the Fe sites 
the LSDA+U approximation was applied~\cite{Anisimov1997} with the on-site Coulomb 
repulsion parameter U=4.5~eV and the intra-atomic Hund's rule exchange 
J$_H$=1 eV.~\cite{Streltsov2011,Hearmon2012}

The Liechtenstein's exchange interaction parameter (LEIP) calculation
procedure~\cite{LEIP1} was used to find the inter-site exchange constants  for the 
Heisenberg model written as
\begin{equation}
\label{Heisenberg}
H = \sum_{ij} J \vec S_i \vec S_j,
\end{equation}
where each site in the summation is counted twice. According to this
method, exchange constants $J$ can be calculated as the second 
derivative of the total energy variation at small spin rotation. This
allows to (1) calculate all $J$ in one magnetic configuration
and (2) check whether a given spin structure corresponds to the
ground state or spins on some of the sites
must be reversed. The later is seen from the sign of $J$, calculated in the LEIP
method: if $J$ is negative (i.e. the second derivative of the total energy
is negative) then the total energy has a minimum for a given magnetic 
order, but if $J$ is positive then one should recalculate the 
exchange constant for a given bond in another spin structure, since the curvature of the
total energy surface and hence the value of $J$ in general 
can be different for minima and maxima. 
\begin{figure}[t!]
 \centering
 \includegraphics[clip=false,width=0.35\textwidth]{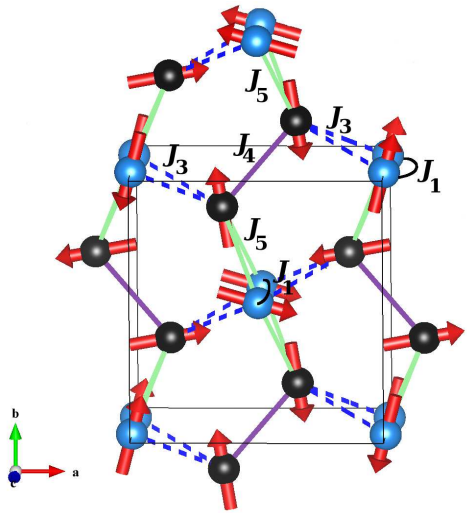}
\caption{\label{magn.str}(color online). The pentagonal magnetic 
lattice of Bi$_2$Fe$_4$O$_9$ (\emph{ab} projection) together with exchange interactions notations. 
The spins on each site are the same (S=5/2), but
the sites have different surrounding: the octahedral Fe$_o$ ions are shown in blue,
the tetrahedral Fe$_t$ ions - in black. The bonds between tetrahedral Fe$_t$ are violet ($J_4$),
while between octahedral Fe$_o$ and tetrahedral Fe$_t$ are green ($J_5$) and blue ($J_3$, dotted line). 
The spin orientation represents the experimentally detected magnetic structure 
from Ref.~\cite{Ressouche-09}.  
In the case of ideal Cairo pentagonal lattice Fe$_t$ ions correspond to the 
threefold-coordinated sites, while Fe$_o$ correspond to the fourfold-coordinated sites. 
Hence, in the notation of Ref.~\onlinecite{Rousochatzakis12}: $J_4 \to J_{33}$ 
and $J_3 =J_5 \to J_{43}$.}
\end{figure}

For the calculation of the
exchange constants between the tetrahedral and octahedral Fe ($J_3$ and $J_5$ in 
Fig.~\ref{magn.str}) we used the magnetic configuration, where the spins
on these bonds are antiferromagnetically coupled, but then the pairs of two
tetrahedral Fe$_t$ ($J_4$) turn out to be ferromagnetically ordered. 
The calculation using the LEIP method shows that the signs for $J_3$ and $J_5$ 
are correct (negative) for this order, but the direction of one of the spins
forming $J_4$ path must be reversed. By checking few other magnetic configurations where spins 
on the Fe$_t$-Fe$_t$ bond were antiferromagnetically ordered we found that
in this case the LEIP method gives negative $J_4$ and its value is the same in these
calculations. The same procedure was repeated for the interplane exchange 
coupling $J_1$ and $J_2$. Since the signs provided by the LEIP procedure does not correspond to 
the usual conventions, in the following the positive (negative) exchange 
constants will mean antiferromagnetic (ferromagnetic) $J$ according to the
Heisenberg model presented in Eq.~\eqref{Heisenberg}.

The crystal structure was taken from Ref.~\onlinecite{Tutov1964} and is shown in 
Fig.~\ref{cryst.str}. The mesh of 144 {\bf k}-points
was used in the course of the self consistency.  

\section{Results}
\begin{figure}[t!]
 \centering
 \includegraphics[clip=false,width=0.5\textwidth]{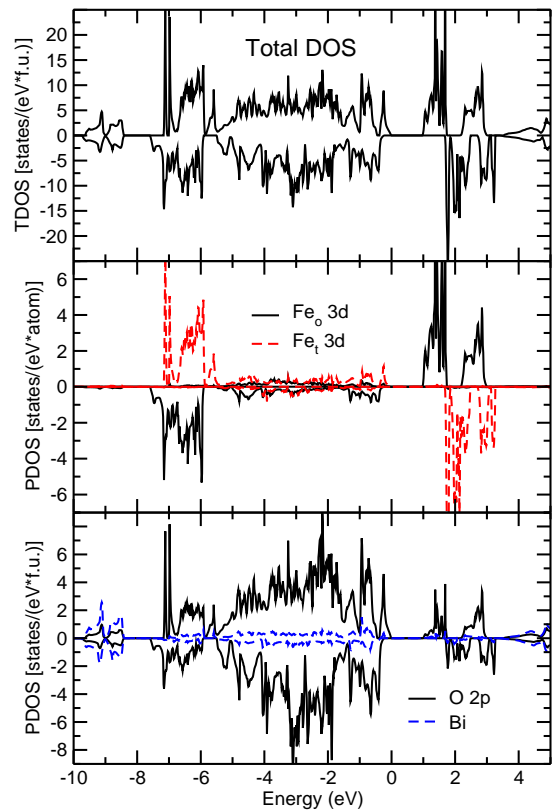}
\caption{\label{DOS}(color online). The total and partial 
density of states plot obtained in the LSDA+U calculation for the
magnetic configuration, where all Fe$_t$ and Fe$_o$ are antiferromagnetically
ordered.
The positive (negative) values correspond to the spin up (down).
The Fermi energy is in zero.}
\end{figure}

The total and partial density of states (DOS) obtained in the LSDA+U calculation for the 
magnetic configuration, where all pairs of Fe$_t$-Fe$_o$ are antiferromagnetically
ordered are presented in Fig.~\ref{DOS}. The DOS obtained for other
magnetic structures is quite similar. The top of the valence
band is mostly defined by the O $2p$ states, while the bottom
of the conduction band is formed by the Fe $3d$ (spin minority) states.
So that Bi$_2$Fe$_4$O$_9$ must be classified as a charge transfer 
insulator.~\cite{ZSA}
The band gap varies from 0.97 eV to 1.28 eV depending on the
magnetic configuration under consideration. The values of the spin moments
are slightly reduced from 5 $\mu_B$ expected for the Fe$^{3+}$ ions with 
$S=5/2$ due to the hybridization effects and equal 3.9-4.0 $\mu_B$.

There are three different types of the exchange coupling in the \emph {ab} 
plane according to our calculations (see Fig.~\ref{magn.str}). The largest is $J_4=73$~K for the pair
of the tetrahedral Fe$_t$. There are also two $J_5 = 23$~K and two 
$J_3 = 36$~K both between the octahedral and tetrahedral Fe ions. 
The main mechanism for all of them is the superexchange via oxygen
ion shared by two FeO$_{6(4)}$ polyhedra. The values of these three exchange constants 
are different because of the quite different geometry of the Fe-O-Fe bonds 
and the ligand polyhedra surrounding each Fe ion.

The exchange constant between two tetrahedral Fe$_t$ ions is the largest, because
of the strong $t_{2g}/t_{2g}$ exchange coupling. The $t_{2g}$ orbitals are 
directed as much as possible to the oxygens in the tetrahedral case and
three $t_{2g}$ orbitals on each Fe$_t$ site take part in a strong
superexchange with the $2p$ orbitals of a common O, via the 180$^{\circ}$ Fe-O-Fe bond.
The direct calculation shows that $J^{t_{2g}/t_{2g}}_4=50$~K, whereas
$J^{t_{2g}/e_g}_4=16$~K and $J^{e_g/e_g}_4=7$~K.

\begin{figure}[t!]
 \centering
 \includegraphics[clip=false,width=0.35\textwidth]{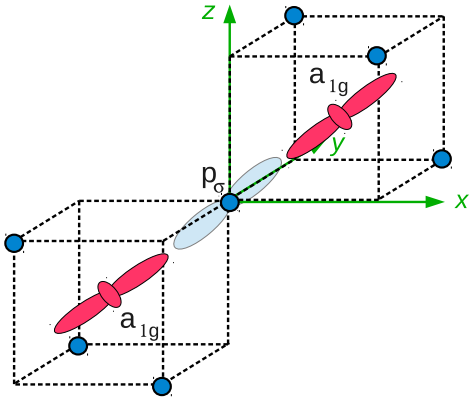}
\caption{\label{tt}(color online). The sketch illustrating the strongest
exchange coupling between $a_{1g}$ orbitals on two tetrahedral Fe$_t$ ions via the 
$p_{\sigma}$ (light blue color) orbital, directed to the centers of 
tetrahedra. Oxygen ions are blue balls.
}
\end{figure}

If the coordinate system is chosen in a way shown in Fig.~\ref{tt}, it's convenient to work 
not with the conventional $p_x$,  $p_y$, and  $p_z$ orbitals, but with 
$p_{\sigma} = (p_x + p_y + p_z)/\sqrt 3$, 
$p_{1} = (p_x - p_y)/\sqrt 2$, and $p_{2} = (p_x + p_y - 2p_z)/\sqrt 6$. 
Then the largest $p-d$ hopping in the tetrahedra will be between
$p_{\sigma}$ and any of the $t_{2g}$ orbitals (different $p-t_{2g}$ hopping 
matrix elements in the case of a regular tetrahedron are calculated in 
Tab.~\ref{Slater-Koster} with the use of the Slater-Koster 
parametrization~\cite{Slater1954}). Hence the largest contribution to the total
exchange interaction between two tetrahedral Fe$_t$ will be
the superexchange via the $p_{\sigma}$ orbital:
\begin{equation}
\label{Jtt}
 J^{t_{2g}/p_{\sigma}/t_{2g}}_{tt} \sim 9 \frac{(t_{p_{\sigma}t_{2g}}^{tet})^2 (t_{p_{\sigma}t_{2g}}^{tet})^2}{U \Delta_{CT}^2},
\end{equation}
where $\Delta_{CT}$ is the charge transfer energy (energy of the excitation 
from the O $2p$ orbitals to the $3d$ shell of a transition metal ion, in our case,
Fe$^{3+}$($d^5$)O$^{2-}$(2$p^6$) $\rightarrow$ Fe$^{2+}$($d^6$) O$^-$(2$p^5$)),~\cite{ZSA} and 
$t_{p_{\sigma}t_{2g}}^{tet}$ is the hopping matrix element between $p_{\sigma}$ and 
one of the $t_{2g}$ orbitals in the FeO$_4$ tetrahedron. Factor 9 comes from
the number of the different $t_{2g}$ orbitals on each site.
In the case of the $t_{2g}/e_g$ exchange interaction this prefactor will
be smaller, so are the hoppings integrals (there will be mostly
$t_{pd\pi}$ hoppings, which are approximately two times
smaller than $t_{pd\sigma}$~\cite{Harrison1999}). 

It is interesting that Eq.~\eqref{Jtt} can be rewritten in a more
useful form if one will use a basis of the trigonal-like~\cite{BiFeO-comment} orbitals also
for the $3d$ states, i.e. $a_{1g} = (d_{xy} + d_{yz} + d_{zx})/\sqrt 3$,
$t_1 = (d_{yz} - d_{zx})/\sqrt 2$, and 
$t_2 = (d_{yz} + d_{zx} - 2d_{xy})/\sqrt 6$. Then all $t_{pd}$
hopping parameters will be zero except $t_{p_{\sigma}a_{1g}}$=$t_{pd\sigma}$,
$t_{p_1 t_1}$=$t_{p_2 t_2}$=$-t_{pd\pi}/\sqrt 3$ and
there will be only two contributions to the exchange between the $t_{2g}$
orbitals coming from the $a_{1g}$ orbitals:
\begin{equation}
 J^{a_{1g}/p_{\sigma}/a_{1g}}_{tt} \sim  \frac{(t_{pd\sigma}^{tet})^4}{U \Delta_{CT}^2},
\end{equation}
and from the $t_1$  and $t_2$ orbitals
\begin{equation}
 J^{t_1/p_{\sigma}/t_1}_{tt} \sim \frac{2(t_{pd\pi}^{tet})^4}{9U \Delta_{CT}^2}.
\end{equation}
Using the estimation of the interatomic matrix elements~\cite{Harrison1999} it's easy
to find that the ratio
\begin{equation}
\frac{ J^{a_{1g}/p_{\sigma}/a_{1g}}_{tt}}
     {J^{t_1/p_{\sigma}/t_2}_{tt}} \sim 100,
\end{equation}
so that one may think that in the case of the regular tetrahedra 
the $t_{2g}/t_{2g}$ exchange with a good precision can be described solely by the superexchange 
between the $a_{1g}$ orbitals via the $p_{\sigma}$ orbital.

\begin{table}
\centering \caption{\label{Slater-Koster} The values of the
$p-d$ hopping matrix elements ($t_{pd}$) between the $d$ and 
$p$ orbitals ($p_{\sigma} = (p_x + p_y + p_z)/\sqrt 3$,
$p_{1} = (p_x - p_y)/\sqrt 2$, and $p_{2} = (p_x + p_y - 2p_z)/\sqrt 6$) in the
case of the regular MeO$_4$ tetrahedron using the Slater-Koster 
parametrization~\cite{Slater1954}, if the coordinate system is chosen
as shown in Fig.~\ref{tt}.}
\vspace{0.2cm}
\begin{tabular}{ccccccc}
\hline
\hline
   & $p_{\sigma}$                   & $p_1$                                &  $p_2$                 \\
\hline
$d_{xy}$ & $\frac 1{\sqrt 3}t_{pd\sigma}$ & 0                              & $\frac {\sqrt 2}3 t_{pd\pi}$ \\
$d_{yz}$ & $\frac 1{\sqrt 3}t_{pd\sigma}$ & $\frac {-1}{\sqrt 6}t_{pd\pi}$ & $\frac {-1}{3\sqrt 2}t_{pd\pi}$ \\
$d_{zx}$ & $\frac 1{\sqrt 3}t_{pd\sigma}$ & $\frac 1{\sqrt 6}t_{pd\pi}$    & $\frac {-1}{3\sqrt 2}t_{pd\pi}$ \\
\hline
\hline
\end{tabular}
\end{table}

Since $J_4 $ is considerably larger than other in-plane exchange couplings
it fixes the directions of the spin moments on two out of three tetrahedral
Fe sites, i.e. makes spins of these Fe ions antiparallel. 
The exchange constants $J_3$  and $J_5$ describing coupling between the 
octahedral Fe$_o$ and tetrahedral Fe$_t$ ions are noticeably smaller than $J_4$ for the 
pair of the tetrahedral Fe$_t$. There are two reasons for that. 

First of all the angle of the Fe$_t$-O-Fe$_o$ bond is far from 180$^{\circ}$ 
(the Fe$_t$-O-Fe$_t$ bond angle is exactly 180$^{\circ}$). If it was $\sim$180$^{\circ}$ then
the superexchange between the Fe$_t$ $t_{2g}$ and Fe$_o$ $e_g$ states via the O $p_{\sigma}$ orbital
would be of order of the $t_{2g}/t_{2g}$ superexchange in the pair of the tetrahedral Fe$_t$ 
ions (the number of active orbitals (two $e_g$ orbitals) of the octahedral Fe$_o$ will be smaller
then in the tetrahedral case (three $t_{2g}$ orbitals), but they will be directed exactly
to the oxygens). However this is not the case. There are two types of the 
tetrahedron-octahedron bonds in Bi$_2$Fe$_4$O$_9$ structure: one with the Fe$_t$-O-Fe$_o$ angle 
$\alpha_1 \sim$120$^{\circ}$ and another with $\alpha_2 \sim$130$^{\circ}$ (see Fig.~\ref{cryst.str}). 
The first one provides
exchange coupling $J_5$, while the second -- $J_3$. Since $\alpha_{1,2}$ are far from both
180$^{\circ}$ and 90$^{\circ}$, the $t_{2g}/t_{2g}$ and $t_{2g}/e_g$ superexchanges 
should be comparable. The direct calculation shows that for the 120$^{\circ}$ bond:
$J_5^{t_{2g}/t_{2g}}$=8~K and $J_5^{t_{2g}/e_{g}}$=8~K, while for the 130$^{\circ}$ bond
$J_3^{t_{2g}/t_{2g}}$=13~K and $J_3^{t_{2g}/e_{g}}$=15~K. Note, that the contribution
coming from the $e_g$ orbitals is surprisingly almost the same for these two exchange
pairs: $J_5^{e_g/e_{g}}$=7~K and $J_3^{e_g/e_{g}}$=8~K.

The second important difference between exchanges in the Fe$_t$--Fe$_t$ and Fe$_t$--Fe$_o$
pairs is in the Fe-O bond distance. In the case of two tetrahedral Fe$_t$ ions both Fe$_t$-O
bond distances are $d(Fe_t-O) = 1.81$ \AA. While for the Fe$_t$--Fe$_o$ pair in the case
of the bond angle $\alpha_1 = 120^{\circ}$ ($J_5$): $d(Fe_t-O) = 1.91$ \AA~and  $d(Fe_o-O) = 2.03$ \AA,
while for $\alpha_2 = 130^{\circ}$ ($J_3$): $d(Fe_t-O) = 1.85$ \AA~and  $d(Fe_o-O) = 1.97$ \AA.
The $Fe_o-O$ bond distances are larger than $Fe_t-O$ since the ionic radius of
the Fe$^{3+}$ is larger in the octahedral coordination than in tetrahedral
($R^{IV}_{HS}(Fe^{3+})=0.49$~\AA~while $R^{VI}_{HS}(Fe^{3+})=0.645$~\AA).~\cite{Shannon-76}

It is rather complicated to find analytically the bond and angle dependence of 
all exchange constants due to the strongly distorted crystal structure and many active 
(magnetically) orbitals in Bi$_2$Fe$_4$O$_9$. We performed such calculations
for the $t_{2g}/e_{g}$ contribution to the 
exchange coupling between the octahedral and tetrahedral Fe ions with the 130$^{\circ}$ 
and 120$^{\circ}$ Fe$_t$-O-Fe$_o$ bonds angles. 
Within the 4th order of the perturbation theory and using approximations
that as it was shown above the Fe$_t$-O hoppings occur only via the 
$p_{\sigma}$ orbital and that they depend only on the Fe$_t$-O bond distance 
one may find that
\begin{equation}
 J^{t_{2g}-e_{g}}_{to} \sim \sum_{i} \frac{(t_{p_{\sigma}a_{1g}}^{tet})^2 
(t_{p_{\sigma}e_g^i}^{oct})^2}{U \Delta_{CT}^2} \sim
 \sum_i C (t_{p_{\sigma}e_g^i}^{oct})^2,
\end{equation}
where $i$ numerates the $e_{g}$ orbitals of the octahedral Fe$_o$. The $t_{p_{\sigma}e_g^i}$
can be estimated using the Slater-Koster coefficients and atomic
positions of the Fe and O ions. Then if one takes into account only
the angle dependence of the hopping matrix elements 
$J^{t_{2g}/e_{g}}_{to}(130^{\circ})/J^{t_{2g}/e_{g}}_{to}(120^{\circ})=1.45$.
The bond length dependence can be found using the Harrison parametrization
of the $pd$ hopping integrals ($t_{pd} \sim \frac 1 {r^{3.5}}$),~\cite{Harrison1999}
which gives $J^{t_{2g}/e_{g}}_{to}(130^{\circ})/J^{t_{2g}/e_{g}}_{to}(120^{\circ})=1.22$.
Taking into account both mechanisms (the angle and bond length dependence)
one finds that this ratio is $\sim 1.8$, which agrees reasonably with
the same ratio, obtained in the LSDA+U calculation, which equals 1.9. 

Calculated exchange constants are in qualitative agreement with the
estimations made in Ref.~\onlinecite{Ressouche-09}. All exchange constants in the
$ab$ plane are antiferromagnetic and the ratio of two tetrahedral-octahedral
exchange constants $J_3/J_5  \approx 1.6$ (2.15 in  Ref.~\onlinecite{Ressouche-09}).
Because of the difference between $J_3$ and $J_5$ Bi$_2$Fe$_4$O$_9$ cannot
be considered as a perfect realization of the Cairo pentagonal 
lattice,~\cite{Ressouche-09} but still the deviations are not so strong, and it makes 
sense to compare our situation with that of the ideal lattice. 
There are only two exchange constants in the perfect
version of this lattice $J_4$ and $J_3 = J_5$. The model study
of the magnetic properties of the ideal Cairo pentagonal lattice shows
that its ground state corresponds to the orthogonal spin order, if
$J_3 / J_4 < \sqrt 2$.~\cite{Rousochatzakis12} 
According to our calculations both $J_5 / J_4 \approx 0.32$ and $J_3 / J_4 \approx 0.49$
are less than $\sqrt 2$, and hence the ground state is also expected
to be described by the orthogonal spin order, exactly as it was observed 
experimentally.~\cite{Ressouche-09}

There are two types of the exchange constants, which couple the octahedral Fe$_o$
ions along the $c$ axis. The first one, $J_1$ (Fe$_o$-Fe$_o$ bond distance 
2.90~\AA), actually has to be considered as a part of the pentagonal lattice (see Fig.~\ref{magn.str}).
This constant is antiferromagnetic and equals $J_1$=10~K, almost a half of one of
the in-plane exchanges ($J_5$). It brings additional (to pentagonal) frustration in
the spin system, since there are four antiferromagnetic triangles 
linked with each pentagon, see Fig.~\ref{magn.str}. The second interplane exchange, 
$J_2$=12~K (Fe$_o$-Fe$_o$ bond distance 3.10~\AA), is antiferromagnetic as well and  
couples different pentagonal planes with each other.

\section{Conclusions}
In the present paper we carried out the band structure calculations
of Bi$_2$Fe$_4$O$_9$ and found that it must be classified as a 
charge transfer insulator. The investigation of the exchange 
constants shows that this compound cannot be considered as a perfect 
realization of the Cairo pentagonal lattice. First of all there 
are two different exchange parameters between the tetrahedral and
octahedral Fe ions. Second the interplanar exchange
coupling additionally frustrates the system. The exchange constants along the $c$ 
axis are not negligibly small 
and exceed 50\% of one of the intraplanar exchange ($J_5$). However, in spite of these
findings Bi$_2$Fe$_4$O$_9$ still demonstrates nearly orthogonal spin 
order below T$_N$~\cite{Ressouche-09} in accordance with the results obtained in 
Ref.~\onlinecite{Rousochatzakis12}, where the study of the
perfect Cairo pentagonal model was performed. This is due to the fact
that both exchange parameters between the tetrahedral and
octahedral Fe ions ($J_3$ and $J_5$) are much smaller than the magnetic coupling
between the tetrahedral Fe sites ($J_4$). Strong $J_4$ makes
the spins on two out of three tetrahedral sites antiparallel, while
the ratio between $J_3$ and $J_5$ define the angles between spin moments 
on the rest one tetrahedral and two octahedral Fe sites.
The microscopic analysis shows that the largest contribution, $\sim$70 \%, 
to $J_4$ comes from the coupling between the $t_{2g}$ orbitals on different sites.
The deviations from the perfect 
Cairo pentagonal model are expected for more subtle characteristics
such as e.g., low energy excitation spectra.

\section{Acknowledgments}
The authors thank Prof. P. Radaelli, who drew our attention on this system.
This work is supported by the Russian Foundation for Basic Research
via  RFFI-13-02-00374, RFFI-13-02-00050, RFFI-12-02-31331, the Ministry of education and science of Russia  
(grants 12.740.11.0026, MK-3443.2013.2, 14.A18.21.0889). The part of the calculations were 
performed on the ``Uran'' cluster of the IMM UB RAS.

\bibliography{library}

\begin{thebibliography}{19}
\bibitem[{BiF()}]{BiFeO-comment}
\bibinfo{note}{These are not true trigonal orbitals, which are defined in the
  chosen coordinate system as $\phi = \frac 1{\sqrt{3}} (xy + e^{i \frac{2\pi}3
  n} xz + e^{-i \frac{2\pi}3 n} yz)$, with $n=0$ corresponding to $a_{1g}$ and
  $n=1,-1$ to $e_g^{\pi}$.}
\end{thebibliography}
\end{document}